\documentclass[11pt,draft]{article}

\title{\bf The Schr$\ddot{o}$dinger equation with Hulth$\acute{e}$n potential plus ring-shaped potential.}
\author{\bf D. Agboola\footnote{E-Mail:tomdavids2k6@yahoo.com}}
\date{Department of Pure and Applied Mathematics,\linebreak Ladoke Akintola University of Technology,Oyo State, Nigeria. \linebreak  P.M.B. 4000}

\begin{document}
\maketitle
\vspace{0.5in}
\noindent{\bf Abtract:}We present the solutions of the Schr$\ddot{o}$dinger equation with the Hulth$\acute{e}$n potential plus ring-shape potential for $\ell\neq 0$ states within the framework of an exponential approximation of the centrifugal potential. \linebreak Solutions to the corresponding angular and radial equations are obtained in terms of special functions using the conventional Nikiforov-Uvarov method. The normalization constant for the Hulth$\acute{e}$n potential is also computed.
\maketitle

\vspace{0.5in}
\vspace{1.2in}
\maketitle

\pagebreak

\noindent{\bf 1. Introduction.}\\\\The search for exact bound-state solutions of wave equations, relativistic or non- relativistic, has been an important research area in quantum mechanics. However, over the past decades, problems involving the multidimensional Schr$\ddot{o}$dinger equation have been addressed by many researchers. For instance, Bateman investigated the relationship between the hydrogen atom and a harmonic oscillator potential in arbitrary dimensions [1]. Recently, the \textsl N-dimensional Pseudoharmonic oscillator was discussed by Agboola \textsl {et al} [2].  The \textsl N-dimensional Kratzer-Fues potential was discussed by Oyewumi [3], while the modified Kratzer-Fues potential plus the ring shape potential in \textsl D-dimensions by the Nikiforov-Uvarov method has also been considered [4].

The Hulth$\acute{e}$n potential is one of the important short-range potentials which behaves like a Coulomb potential for small values of  $r$ and decreases exponentially for large values of \textsl r. The Hulth$\acute{e}$n potential has received extensive study in both relativistic and non-relativistic quantum mechanics [5-9]. Unfortunately, quantum mechanical equations with the Hulth$\acute{e}$n potential can be solved analytically only for the \textsl s-states [5, 11, 12]. However, some interesting research papers [13-22] have appeared to study the $\ell-state$ solutions of Hulth$\acute{e}$n-type potentials. Recently, an extension of this study to a multidimensional case was presented [23]. The main idea of the investigation relies on using an exponential approximation for the centrifugal term.

Recently, Chen and Dong [24] found a new ring-shaped (non-central) potential and obtained the exact solution of the Schr$\ddot{o}$dinger equation for the Coulomb potential plus this new ring-shaped potential which has possible applications to ring-shaped organic molecules like cyclic polyenes and benzene. Also, Cheng and Dai [25], proposed a new potential consisting from the modified Kratzer potential [26] plus the new proposed ring-shaped potential. They have presented the energy eigenvalues for this proposed exactly-solvable non-central potential. Very recently, the \textsl D-dimensional case of the potential has been studied by Ikhdair[] using the Nikiforov-Uvarov method [27].

It is therefore the aim of this paper to present the approximate solutions of the Schr$\ddot{o}$dinger equation with the Hulth$\acute{e}$n potential plus ring-shape potential for $\ell\neq 0$   states using the conventional Nikiforov-Uvarov method.\\ 

\pagebreak



\noindent{\bf 2. The  Schr$\ddot{o}$dinger equation for the  Hulth$\acute{e}$n plus ring-shape potential.}\\\\
The motion of a particle of mass $\mu$ in a spherically symmetric potential is described in spherical coordinate as
$$-\frac{\hbar^2}{2\mu}\left[\frac{1}{r^2}\frac{\partial}{\partial r}\left(r^2\frac{\partial}{\partial r}\right)+\frac{1}{r^2}\left(\frac{1}{\sin\theta}\frac{\partial}{\partial \theta}\left(\sin\theta\frac{\partial}{\partial\theta}\right)+\frac{1}{\sin^2\theta}\frac{\partial^2}{\partial\varphi^2}\right)+V(r)\right]\Psi_{n\ell m}(r,\theta,\varphi)$$
$$\hspace{4.5in} =E\Psi_{n\ell m}(r,\theta,\varphi) \eqno{(1)}$$ 
The  Hulth$\acute{e}$n potential [5, 6, 7, 23] plus the ring-shape potential [4, 24, 25] in \textsl  is given as
$$V(r,\theta)=-Z\alpha\frac{e^{-\alpha r}}{1-e^{-\alpha r}}+ \beta\frac{\cos^2\theta}{r^2 \sin^2\theta}   \eqno {(2)} $$
where $\alpha$ is the screening parameter, $\beta$ is a positive real constant and  \textsl Z is a constant which is identified with the atomic number when the potential is used for atomic phenomena.\\  
If we substitute Eq.(2) into (1) and seperate the variable as follows
$$\Psi_{n\ell m}(r,\theta,\varphi)=R_{n\ell}(r)Y_\ell^m(\theta,\varphi)  \eqno{(3)}$$
where
$$R_{n\ell}(r)=r^{-1}U_{n\ell}(r)  \eqno{(4)}$$
and  
$$Y_\ell^m(\theta,\varphi)=H(\theta)\Phi(\varphi),  \eqno{(5)}$$
then, we have the following sets of second-order differential equations: 
$$U_{n\ell}^{\prime\prime}(r)+\left[\frac{2\mu}{\hbar^2}\left(E+Z\alpha\frac{e^{-\alpha r}}{1-e^{-\alpha r}}\right)-\frac{\ell(\ell+1)}{r^2}\right]U_{n\ell}(r)=0, \eqno{(6)}$$\\
$$\left[\frac{1}{\sin\theta}\frac{d}{d\theta}\left(\sin\theta\frac{d}{d\theta}\right)-\frac{m^2}{\sin^2\theta}-\frac{2\mu\beta}{\hbar^2}\frac{\cos^2\theta}{ \sin^2\theta}+\ell(\ell+1)\right]H(\theta)=0  \eqno{(7)}$$\\
and\\ 
$$\frac{d^2\Phi(\varphi)}{d\varphi^2}+m^2\Phi(\varphi)=0  \eqno{(8)}$$
where we have introduced the seperation constants $m$ and $\nu=\ell(\ell+1)$ and $\ell$ is the orbital angular momentum quantum number.

The solution to Eq.(8) is periodic and must satisfies the periodic boundary condition
$$\Phi(\varphi+2\pi)=\Phi(\varphi)  \eqno{(9)}$$
from which we obtain the solution 
$$\Phi_m(\varphi)=\frac{1}{\sqrt{2\pi}}exp(\pm im\varphi),\hspace{0.3in} m=0,1,2,...  \eqno{(10)}$$
We are then left with the solutions of Eqs.(6) and (7) which we present in the later sections.\\

\noindent {\bf 3. The Nikiforov-Uvarov method.}\\\\
In this section, we give a brief description of the conventional Nikiforov-Uvarov method. A more detailed description of the method can be obtained the following references [27].With an appropriate transformation\linebreak $s=s(r)$,the one dimensional Schr$\ddot{o}$dinger equation can be reduced to a generalized equation of hypergeometric type which can be written as follows:
$$\psi^{\prime\prime}(s)+ \frac{\tilde{\tau}(s)}{\sigma(s)}\psi^\prime(s)+ \frac{\tilde{\sigma}(s)}{\sigma^2(s)}\psi(s)=0  \eqno{(11)}$$ 
Where $\sigma(s)$and $\tilde{\sigma}(s)$ are polynomials, at most second-degree, and $\tilde{\tau}(s)$is at most a first-order polynomial. To find particular solution of Eq. (14) by separation of variables, if one deals with
$$\psi(s)=\phi(s)y_n(s),  \eqno{(12)}$$
Eq.(11)becomes
$$\sigma(s)y^{\prime\prime}_n+\tau(s)y^\prime_n +\lambda y_n =0  \eqno{(13)}$$
where
$$\sigma(s)= \pi(s)\frac{\phi(s)}{\phi^\prime(s)},  \eqno{(14)}$$,
$$\tau(s)=\tilde{\tau}(s)+2\pi(s) ,  \tau^\prime(s)<0,  \eqno{(15)}$$\\
$$\pi(s)=\frac{\sigma^\prime-\tilde{\tau}}{2}\pm \sqrt{\left(\frac{\sigma^\prime-\tilde{\tau}}{2}\right)^2-\tilde{\sigma}+t\sigma},  \eqno{(16)}$$
and 
$$\lambda=t+\pi^\prime(s).  \eqno{(17)}$$
The polynomial $\tau(s)$ with the parameter $s$ and prime factors show the differentials at first degree be negative. However,determination of parameter $t$ is the essential point in the calculation of $\pi(s)$. It is simply defined by setting the discriminate of the square root to zero [27]. Therefore, one gets a general quadratic equation for $t$.The values of $t$ can be used for calculation of energy eigenvalues using the following equation
$$\lambda=t+\pi^\prime(s)=-n\tau^\prime(s)-\frac{n(n-1)}{2}\sigma^{\prime\prime}(s).   \eqno{(18)}$$
Furthermore, the other part $y_n(s)$ of the wave function in Eq. (12) is the hypergeometric-type function whose polynomial solutions are given by Rodrigues relation: 
$$y_n(s)=\frac{B_n}{\rho(s)}\frac{d^n}{ds^n}[\sigma^n(s)\rho(s)]  \eqno{(19)}$$ 
where $B_n$ is a normalizing constant and the weight function $\rho(s)$ must satisfy the condition [10]
$$(\sigma\rho)^\prime =\tau\rho.   \eqno{(20)}$$\\

\noindent{\bf 4.0 Solution to the angular equation.}\\\\
In order to solve Eq.(6) using the Nikiforov-Uvarov method, we write it as follows
$$\frac{d^2H(\theta)}{d\theta^2}+\frac{\cos\theta}{\sin\theta}\frac{dH(\theta)}{d\theta}+\left[\ell(\ell+1)-\frac{m^2+(2\mu\beta\slash \hbar^2)\cos\theta}{\sin^2\theta}\right]H(\theta)=0  \eqno{(21)}$$
Introducing a new variable $s=\cos\theta$, Eq. (21) becomes an associated Legendre differential equation [4, 25, 28]  
$$\frac{d^2H(s)}{ds^2}+\frac{2s}{1-s^2}\frac{dH(s)}{ds}+\frac{\eta(1-s^2)-\kappa^2}{(1-s^2)^2}H(s)=0  \eqno{(22)}$$
where
$$\eta=\ell^\prime(\ell^\prime+1)=\ell(\ell+1)+ \frac{2\mu\beta}{\hbar^2}\hspace{0.1in} and\hspace{0.1in} \kappa^2=m^2+ \frac{2\mu\beta}{\hbar^2}  \eqno{(23)}$$
Comparing Eqs. (11) and (23) we obtained the following
$$\tilde{\tau}(s)=-2s,\hspace{0.2in} \sigma(s)=1-s^2 \hspace{0.1in} and \hspace{0.1in} \tilde{\sigma}(s)=-\eta s^2+\eta-\kappa^2   \eqno{(24)} $$ 
Inserting the above expressions in Eq. (16) we obtain the following functions: 
$$\pi(s)=\pm\sqrt{(\eta-t)s^2+t-\eta+\kappa^2}  \eqno{(25)}$$ 
Equating the discriminate of the expression under the square root to zero to get the possible values of  $t$, we then have the corresponding values of $\pi(s)$ as follows:
$$\pi(s)= \left\{\begin{array}{rll}
 \kappa s & \mbox{for} & t=\eta-\kappa^2\\
 -\kappa s & \mbox{for} & t=\eta-\kappa^2\\
\kappa & \mbox{for} & t=\eta\\
 -\kappa & \mbox{for} & t=\eta\end{array}\right. \eqno{(26)}$$
With the condition that $\tau^\prime(s)<0$, we select the solution $\pi(s)=-\kappa s$ for $t=\eta-\kappa^2$.This yields $$\tau(s)=-2(1+\kappa)s.  \eqno{(27)}$$ Using Eq.(18), we have the following $$\lambda=\eta-\kappa(1+\kappa)=2n(1+\kappa)+n(n-1)  \eqno{(28)}$$
Eq.(28)with the definition $\eta=\ell^\prime(\ell^\prime+1)$ yields the new angular momentum $\ell^\prime$ as 
$$\ell^\prime= n+\kappa=n+\sqrt{m^2+ \frac{2\mu\beta}{\hbar^2}}.  \eqno{(29)}$$
Moreover, using Eqs.(14),(19)and (20),it is easy to obtain the following
$$\phi(s)=(1-s^2)^{\kappa/2}  \eqno{(30)}$$
and
$$y_n=B_n(1-s^2)^{-\kappa}\frac{d^n}{ds^n}\left[(1-s^2)^{n+k}\right]  \eqno{(31)}$$  
such that with the substitution $s=cos\theta$, we have the angular solution to be
$$H_\kappa(\theta)=N_{\ell^\prime\kappa}\sin^\kappa(\theta)P_n^{(\kappa,\kappa)}(\cos\theta)  \eqno{(32)}$$ where $N_{\ell^\prime\kappa}$ is the normalization canstant given as $$N_{\ell^\prime\kappa}=\sqrt{\frac{(2\ell^\prime+1)(\ell^\prime-\kappa)!}{2(\ell^\prime+\kappa)!}}  \eqno{(32)}$$ and $n=\ell^\prime+\kappa$.\\\\

\pagebreak
\noindent{\bf 4.1 Solution to the radial equation.}\\\\
In this secion, we obtain the solution to radial equation (6) using the Nikiforov-Uvarov method.First, we note that Eq.(6) is similar to the one dimensional Schr$\ddot{o}$dinger equation for the Hulth$\acute{e}$n potential, expect for the addition of the term $\frac{\ell(\ell+N-2)}{r^2}$,which is well known as the centrifugal term.To solve Eq.(6),one can consider the approximation of the centrifugal term which is valid for a small value of $\alpha $ 
$$\frac{1}{r^2}\approx\frac{\alpha^2e^{-\alpha r}}{(1-e^{-\alpha r})^2}.  \eqno{(33)}$$
With the use of Eq.(33) and the transformation $s=e^{-\alpha r}$, Eq.(6) becomes
$$U_{n\ell}^{\prime\prime}(s)+\frac{(1-s)}{s(1-s)}U_{n\ell}^\prime(s)+\frac{1}{[s(1-s)]^2}[(-\epsilon^2-\delta)s^2+(2\epsilon^2+\delta-\gamma)s-\epsilon^2]U_{n\ell}(s)=0  \eqno{(34)}$$
where
$$-\epsilon^2=\frac{2\mu E}{\alpha^2 \hbar^2},\hspace{.2in} \delta=\frac{2Z\mu}{\alpha\hbar^2},\hspace{.2in} and \hspace{.2in} \gamma=\ell(\ell+1)  \eqno{(35)}$$
By comparing Eqs.(11) and (34), we can define the following
$$\tilde{\tau}(s)=1-s,\hspace{.2in} \sigma(s)=s(1-s)\hspace{.2in} and \hspace{.2in}\tilde{\sigma}(s)=(-\epsilon^2-\delta)s^2+(2\epsilon^2+\delta-\gamma)s-\epsilon^2  \eqno{(36)}$$
Inserting these expressions into Eq.(16), we have
$$\pi(s)=-\frac{s}{2}\pm \frac{1}{2}\sqrt{[1+4(\epsilon^2+\delta-t)]s^2-4(2\epsilon^2+\delta-\gamma+t)s+4\epsilon^2}  \eqno{(37)}$$ 
The constant parameter $t$ can be found in a similar fashion with that of the angular equation. Thus, the two possible functions for each $t$ are given as\\
$$\pi(r)=-\frac{s}{2}\pm \left\{\begin{array}{rll}
\frac {1}{2}\left [\left(2\epsilon -\sqrt{1+4\gamma}\right)s-2\epsilon\right] & \mbox{for} & t=\gamma-\delta+\epsilon\sqrt{1+4\gamma}\\\\
\frac {1}{2}\left[\left(2\epsilon +\sqrt{1+4\gamma}\right)s-2\epsilon\right] & \mbox{for} & t=\gamma-\delta-\epsilon\sqrt{1+4\gamma}\end{array}\right. \eqno{(38)}$$\\
However, for the polynomial $\tau(s)=\tilde{\tau}(s)+2\pi(s)$ to have a negative derivative, we can select the physically valid solution to be$$\pi(s)=-\frac{s}{2}-\frac {1}{2}\left[\left(2\epsilon +\sqrt{1+4\gamma}\right)s-2\epsilon\right]  \eqno{(39)}$$ for $ t=\gamma-\delta-\epsilon\sqrt{1+4\gamma}$ such that$$\tau(s)=1-2s-\left[\left(2\epsilon +\sqrt{1+4\gamma}\right)s-2\epsilon\right]. \eqno{(40)}$$
Also,by Eq.(16) we can define
$$\lambda=\gamma-\delta-\frac{1}{2}(1+2\epsilon)\left(1+\sqrt{1+4\gamma}\right)=n\left[1+2\epsilon+n+\sqrt{1+4\gamma}\right]  \eqno{(41)}$$for $n=1,2,...$
Thus, with the aid of Eqs.(35)and (41),after simple manipulations, we have the energy eigenvalues as
$$E_n=-\frac{\hbar^2}{2\mu}\left[\frac{(Z\mu/\hbar^2)}{n+\ell+1}-\frac{n+\ell+1}{2}\alpha\right]^2  \eqno{(42)}$$ which is in agreement with previous works [6,28].

To obtain the wave function, we substitute $\pi(s)$ and $\sigma(s)$ into Eq.(14), and solving the first order differential  equation to have
$$\phi(s)=s^\epsilon(1-s)^{\ell+1}.  \eqno{(43)}$$
Also by Eq.(20), the weight function $\rho(s)$ can be obtained as 
$$\rho (s)=s^{2\epsilon}(1-s)^{2\ell+1}  \eqno{(44)}$$
Substituting Eq.(42)into the Rodrigues relation (19), we have
$$y_n(s)=B_ns^{-2\epsilon}(1-s)^{-(2\ell+1)}\frac{d^n}{ds^n}\left[s^{n+2\epsilon}(1-s)^{n+2\ell+1}\right]. \eqno{(45)}$$ 
Therefore, we can write the wave function $U_{n\ell}(s)$ as 
$$U_{n\ell}(s)=C_ns^\epsilon(1-s)^{\ell+1}P^{(2\epsilon,2\ell+1)}_n(1-2s)  \eqno{(46)}$$
where $C_n$ is the normalization constant, and we have used the definition of the Jacobi polynomials[30],given as
$$P^{(a,b)}_n(s)=\frac{(-1)^n}{n!2^n(1-s)^a(1+s)^b}\frac{d^n}{ds^n}\left[(1-s)^{a+n}(1+s)^{b+n}\right].  \eqno{(47)}$$
To compute the normalization constant $C_n$, it is easy to show with the use of Eq.(4) that $$\int^\infty_0|R_{n\ell}(r)|^2r^{2}dr=\int^\infty_0|U_{n\ell}(r)|^2dr=\int^1_0|U_{n\ell}(s)|^2\frac{ds}{\alpha s}=1  \eqno{(48)}$$
where we have also used the substitution $s=e^{-\alpha r}$. Putting Eq.(46) into Eq.(48) and using the following definition of the Jacobi polynomial[30]
$$P^{(a,b)}_n(s)=\frac{\Gamma(n+a+1)}{n!\Gamma(1+a)} \ _2F_1\left(-n,a+b+n+1;1+a;\frac{1-s}{2}\right),  \eqno{(49)}$$ we arrived at 
$$C_n^2N^\epsilon_n\int_0^1s^{2\epsilon-1}(1-s)^{2\ell+2}\ _2F_1\left(-n,2\epsilon+2\ell+n+2;1+2\epsilon;s\right)ds=\alpha \eqno{(50)}$$ where$$N^\epsilon_n=\left[\frac{\Gamma(2\epsilon+n+1)}{n!\Gamma(2\epsilon+1)}\right]^2  \eqno{(51)}$$ and $_2F_1$ is the hypergeometric function. Using the following series representation of the hypergeometric fucntion
$$_pF_q(a_1,...,a_p;c_1,...,c_q;s)=\sum_{n=0}^\infty\frac{(a_1)_n...(a_p)_n}{(c_1)_n...(c_q)_n}\frac{s^n}{n!}  \eqno{(52)}$$we have
$$C_n^2N^\epsilon_n\sum^n_{k=0}\sum^n_{j=0}\frac{(-n)_k(n+2\epsilon+2\ell+2)_k}{(1+2\epsilon)_k k!}\frac{(-n)_j(n+2\epsilon+2\ell+2)_j}{(1+2\epsilon)_j j!}\int_0^1s^{2\epsilon+k+j-1}(1-s)^{2\ell+2} ds=\alpha .\eqno{(53)}$$
Hence, by the definition of the Beta function,Eq.(53)becomes
$$C_n^2N^\epsilon_n\sum^n_{k=0}\sum^n_{j=0}\frac{(-n)_k(n+2\epsilon+2\ell+2)_k}{(1+2\epsilon)_k k!}\frac{(-n)_j(n+2\epsilon+2\ell+2)_j}{(1+2\epsilon)_j j!} B(2\epsilon+k+j,2\ell+3)=\alpha.  \eqno{(54)}$$
Using the relations $B(x,y)=\frac{\Gamma(x)\Gamma(y)}{\Gamma(x+y)}$ and the Pochhammer symbol \linebreak $(a)_n=\frac{\Gamma(a+n)}{\Gamma(a)}$, Eq.(54) can be written as 
$$C_n^2N^\epsilon_n\sum^n_{k=0}\frac{(-n)_k(2\epsilon)_k(n+2\epsilon+2\ell+2)_k}{(1+2\epsilon+2\ell+2)_k (1+2\epsilon)_k k!}\sum^n_{j=0}\frac{(-n)_j(2\epsilon+k)_j(n+2\epsilon+v)_j}{(1+2\epsilon+v+k)_j (1+2\epsilon)_j j!}=\frac{\alpha}{B(2\epsilon,2\ell+3)}  \eqno{(55)}$$

\noindent Eq.(55) can be used to compute the normalization constants for $n=0,1,2,...$ In particular for the gound state, $n=0$, we have
$$C_0=\sqrt{\frac{\alpha}{B(2\epsilon,2\ell+3)}}  \eqno{(56)}$$

Finally, we can now write the complete orthonormalized energy eigenfunction of the Hulth$\acute{e}$n potential plus a ring-shape potential. 
$$\Psi_{n\ell m}(r,\theta,\varphi)=r^{-1}U_{n\ell}(s)H(\theta)\Phi(\varphi)  \eqno{(57)}$$
where $U_{n\ell}(s)$,$H(\theta)$ and $\Phi(\varphi)$ are give in Eqs.(46),(32) and(10) respectively and $s=e^{-\alpha r}$\\\\
\noindent{\bf 6. Conclusions.}\\\\ 
In this paper, the Schr$\ddot{o}$dinger equation was solved for its approximate bound-states with a  Hulth$\acute{e}$n potential plus a ring-shaped potential for $\ell\neq 0$ by the conventional Nikiforov-Uvarov method. Solutions to the angular and radial equations were obtained using the Nikiforov-Uvarov method and the eigenvalues obtained were found to be in agreement with those obtained using the numerical integration method[31] and the PT-symmetric quantum mechanics[28].The corresponding eigenfunctions were worked out in terms of the Jacobi polynomial and the normalization constant was also computed in terms of hypergeometric series.This research is an extension of the works presented by [24]. 
 
However,it is important to note that the approximation (33) is only valid for a small value of $\alpha $; and as $\alpha\rightarrow 0$, the results obtained approach those of the Coulombic potential.

Finally, it is worth noting that the approximate solution obtained in the newly proposed form of potential (2) may have some significant applications in the study of quantum mechanical systems in both atomic and molecular physics.

\pagebreak
                                     
\end{document}